\documentclass[english]{article}%
\usepackage[T1]{fontenc}
\usepackage[latin1]{inputenc}
\usepackage{amsmath}
\usepackage{babel}
\usepackage{graphicx}
\usepackage{amsfonts}
\usepackage{amssymb}
\usepackage[a4paper]{geometry}%
\makeatletter
\newtheorem{theorem}{Theorem}

\newtheorem{acknowledgement}[theorem]{Acknowledgement}

\makeatother
\begin{document}

\title{Charged particles in crossed and longitudinal electromagnetic fields and beam guides.}
\author{V.G. Bagrov\thanks{Tomsk State University; Institute of High Current
Electronics of the Siberian Branch of the Russian Academy of Sciences. E-mail:
bagrov@phys.tsu.ru}, M.C. Baldiotti\thanks{E-mail: baldiott@fma.if.usp.br},
and D.M. Gitman\thanks{E-mail: gitman@dfn.if.usp.br}\\Instituto de Física, Universidade de São Paulo,\\Caixa Postal 66318-CEP, 05315-970 São Paulo, S.P., Brazil}
\date{}
\maketitle

\begin{abstract}
We consider a class of electromagnetic fields that contains crossed fields
combined with longitudinal electric and magnetic fields. We study the motion
of a classical particle (solutions of the Lorentz equations) in such fields.
Then, we present an analysis that allows one to decide which fields from the
class act as a beam guide for charged particles, and we find some
time-independent and time-dependent configurations with beam guiding
properties. We demonstrate that the Klein-Gordon and Dirac equations with all
the fields from the class can be solved exactly. We study these solutions,
which were not known before, and prove that they form complete and orthogonal
sets of functions.

\end{abstract}

\section{Introduction}

Relativistic wave equations (Dirac and Klein--Gordon) provide a basis for
relativistic quantum mechanics and QED of spinor and scalar particles. In
relativistic quantum mechanics, solutions of relativistic wave equations are
referred to as one-particle wave functions of fermions and bosons in external
electromagnetic fields. In QED, such solutions permit the development of the
perturbation expansion known as the Furry picture, which incorporates the
interaction with the external field exactly, while treating the interaction
with the quantized electromagnetic field perturbatively
\cite{Schwe61,GreMuR85,BagGi90,FraGiS91,Grein97}. The most important exact
solutions of the Klein--Gordon and Dirac equations are: solutions with the
Coulomb field, which allow one to construct the relativistic theory of atomic
spectra \cite{BetSa57}, solutions with a uniform magnetic field, which provide
the basis of synchrotron radiation theory \cite{synchrotron}, and solutions in
the field of a plane wave, which are widely used for calculations of quantum
effects involving electrons and other elementary particles in laser beams
\cite{laser}. Another physically important class of field configurations (for
solving the relativistic wave equations) is a superposition of crossed fields
and longitudinal fields. Solutions of relativistic equations with fields of
this type were first studied by Redmond \cite{Red65}. The Redmond
configuration is a plane-wave combined with a constant longitudinal magnetic
field. The corresponding solutions have wide spread applications, for example,
in plasma physics \cite{BroBu62} and cyclotron resonance \cite{RobBu64}. In
the works \cite{35,78,117} exact solutions of the relativistic wave equations
with a generalized Redmond configuration (Redmond field plus longitudinal
electric fields) were found and used to calculate different quantum effects.
In \cite{Bir04} the author has presented another generalization of a crossed
field, which is a particular (the simplest) case of a vortex field
\cite{Bir03,Bir04b} (electromagnetic waves with vortices play a central role
in singular optics \cite{2a}). He studied exact solutions of relativistic wave
equations in such a field and he has discovered that it can be used to create
a beam guide for charged particles.

In the present article we represent and study new solutions of the
Klein--Gordon and Dirac equations with a new class of fields, which is a
combination of crossed and longitudinal electromagnetic fields. For the
crossed fields $E_{z}=H_{z}=0,\;E_{x}=H_{y}$ and $E_{y}=-H_{x}$, and they
depend on the time $t$ and on the coordinate $z$ via a light-cone variable
$\xi=ct-z$. In the general case, the amplitudes of the crossed fields can also
contain a linear $\xi$-dependent combination of the coordinates $x,y$. Thus,
we can interpret the crossed fields as plane-waves with amplitudes linearly
dependent on the coordinates $x,y$. One ought to say that this combination of
crossed and longitudinal fields form a class which is described by several
arbitrary $\xi$-dependent functions. This combination of electromagnetic
fields is physically interesting, since some configurations act as beam guides
for charged particles in a similar fashion to which the vortex field acts in
the work \cite{Bir04}. The aforementioned vortex field is a particular case of
our beam guiding configurations. It is interesting to stress that all other
beam-guiding field configurations do not belong to the vortex field class.

The article is organized as follows: first, we describe potentials for the
above mentioned combination of electromagnetic fields and we study classical
particle motion (that is, solutions of the Lorentz equations) in such fields.
Then, we present an analysis that allows one to decide which fields from the
combination act as a beam guide for charged particles. We find some
time-independent and time-dependent configurations with beam guiding
properties. Finally, we study solutions of the Klein-Gordon and Dirac
equations containing all the fields from the combination and we prove that
these solutions form complete and orthogonal sets of functions. In the
Appendix, we place some technical results.

The electromagnetic fields we consider are defined by the following
potentials\footnote{The four-dimensional coordinates of a particle are denoted
as $x^{\mu}=\left(  x^{0}=ct,\, x^{1}=x,\, x^{2}=y,x^{3}=\, z\right)  $,
$\mu=0,1,2,3$, where $c$ is the speed of light. Contravariant and covariant
four-vectors are often represented in the form%
\begin{align*}
a^{\mu}  &  = \left(  a^{0},a^{i}\right)  =\left(  a^{0},\mathbf{a}\right)
,\;\mathbf{a=}\left(  a^{i}\right)  \,,\; a^{1}=a_{x},\; a^{2}=a_{y},\;
a^{3}=a_{z}\,,\\
a_{\mu}  &  = \eta_{\mu\nu}a^{\nu}\,,\; a^{\mu}=\eta^{\mu\nu}a_{\nu}\,.
\end{align*}
Three-vectors are indicated by boldface letters.}:%
\begin{align}
&  A^{0}=\frac{1}{2}\left[  \mathcal{G}\left(  \xi\right)  -A\right]  ,\;
A^{1}=A_{x}=-\mathcal{F}_{1}\left(  \xi\right)  -\mathcal{H}\left(
\xi\right)  y\,,\nonumber\\
&  A^{2}=A_{y}=-\mathcal{F}_{2}\left(  \xi\right)  +\mathcal{H}\left(
\xi\right)  x,\; A^{3}=A_{z}=-\frac{1}{2}\left[  \mathcal{G}\left(
\xi\right)  +A\right]  \,, \label{1a}%
\end{align}
where%
\[
A={R}_{11}\left(  \xi\right)  x^{2}+2{R}_{12}\left(  \xi\right)  xy+{R}%
_{22}\left(  \xi\right)  y^{2},\;\xi=x^{0}-z=ct-z\,,
\]
and $\mathcal{G}\left(  \xi\right)  $, $\mathcal{H}\left(  \xi\right)  $,
$\mathcal{F}_{i}\left(  \xi\right)  $, $R_{i\, j}\left(  \xi\right)  =R_{j\,
i}\left(  \xi\right)  ,\, i,j=1,\,2$, are arbitrary functions of $\xi$. The
corresponding electromagnetic fields have the form%
\begin{align}
&  E_{x}=H_{y}=\mathcal{F}_{1}^{\prime}\left(  \xi\right)  +{R}_{11}\left(
\xi\right)  x+\left[  {R}_{12}\left(  \xi\right)  +\mathcal{H}^{\prime}\left(
\xi\right)  \right]  y,\; E_{z}=\mathcal{G}^{\prime}\left(  \xi\right)
~,\nonumber\\
&  E_{y}=-H_{x}=\mathcal{F}_{2}^{\prime}\left(  \xi\right)  +\left[  {R}%
_{12}\left(  \xi\right)  -\mathcal{H}^{\prime}\left(  \xi\right)  \right]
x+{R}_{22}\left(  \xi\right)  y,\; H_{z}=2\mathcal{H}\left(  \xi\right)  ~.
\label{2a}%
\end{align}
They consist of crossed fields and longitudinal electric and magnetic fields
propagating along the $z$-axis. In the general case, amplitudes of the crossed
fields depend linearly on the coordinates $x,y$.

The Maxwell current determined by the field (\ref{2a}) has the form%
\begin{equation}
j^{\mu}=\frac{c}{4\pi}\left(  \rho,0,0,\rho\right)  ,\;\rho=\rho\left(
\xi\right)  =R_{11}\left(  \xi\right)  +R_{22}\left(  \xi\right)
-\mathcal{G}^{\prime\prime}\left(  \xi\right)  ~. \label{3}%
\end{equation}

\section{Classical motion}

Let us first examine the classical Lorentz equations%
\begin{equation}
m_{0}c^{2}\dot{u}^{0}=e\left(  \mathbf{uE}\right)  ,\; m_{0}c^{2}%
\mathbf{\dot{u}}=e\mathbf{E}u^{0}+e\left[  \mathbf{u}\times\mathbf{H}\right]
,\;\left(  u^{0}\right)  ^{2}-\mathbf{u}^{2}=1~, \label{4}%
\end{equation}
where%
\[
u^{\mu}=\frac{dx^{\mu}}{ds}=\dot{x}^{\mu}=\left(  u^{0},\mathbf{u}\right)  ,\;
ds^{2}=\eta_{\mu\,\nu}dx^{\mu}dx^{\nu},\;\eta_{\mu\,\nu}=\mathrm{diag\,}%
\left(  1,-1,-1,-1\right)  ~.
\]
And the Hamilton--Jacobi equation is%
\begin{equation}
\left(  \partial\,_{0}\mathcal{S}+\frac{e}{c}A^{0}\right)  ^{2}-\left(
\nabla\mathcal{S}-\frac{e}{c}\mathbf{A}\right)  ^{2}-m_{0}^{2}c^{2}=0~,
\label{5}%
\end{equation}
where $\mathcal{S}$ is the classical action. From equations (\ref{4})
obviously follow the equations for the kinetic momenta $P^{\mu}=m_{0}c\,
u^{\mu}=\left(  P^{0},\mathbf{P}\right)  :$
\begin{equation}
m_{0}c^{2}{\dot{P}}^{0}=e\left(  \mathbf{PE}\right)  ,\; m_{0}c^{2}%
\mathbf{\dot{P}}=eP^{0}\mathbf{E}+e\left[  \mathbf{P}\times\mathbf{H}\right]
,\;\left(  P^{0}\right)  ^{2}-\mathbf{P}^{2}=m_{0}^{2}c^{2}~. \label{6}%
\end{equation}
In particular, from (\ref{6}), with allowance made for (\ref{2a}), we easily
obtain%
\begin{equation}
m_{0}c^{2}\,{\dot{P}}_{z}=e\left(  \mathbf{PE}\right)  +e\left(  P^{0}%
-P_{z}\right)  E_{z}~. \label{7}%
\end{equation}
Let us introduce the generalized momenta $p_{\mu}$ according to the well-known
relations%
\begin{equation}
P_{\mu}=p_{\mu}-\frac{e}{c}A_{\mu},\; p_{\mu}=-\partial_{\mu}\mathcal{S}~.
\label{8}%
\end{equation}
One can easily prove that the quantity%
\begin{equation}
\Lambda=p^{0}-p_{z} \label{9}%
\end{equation}
is an integral of motion. Indeed, (\ref{8}) implies
\begin{equation}
\Lambda=p^{0}-p_{z}=P\,^{0}-P_{z}+\frac{e}{c}\mathcal{G}\left(  \xi\right)  ~.
\label{10}%
\end{equation}
Hence, we obtain%
\begin{equation}
\dot{\Lambda}=\dot{P}\,^{0}-\dot{P}_{z}+\frac{e}{c}\mathcal{G^{\prime}}\left(
\xi\right)  \dot{\xi}=\dot{P}\,^{0}-\dot{P}_{z}+\frac{e}{c}E_{z}\dot{\xi}~.
\label{11}%
\end{equation}
Taking into account the obvious relation%
\begin{equation}
\dot{\xi}=\dot{x}^{0}-\dot{z}=u^{0}-u_{z}=\frac{P\,^{0}-P_{z}}{m_{0}c}
\label{12}%
\end{equation}
and the equations (\ref{6}) and (\ref{7}), we find that (\ref{11}) implies
$\dot{\Lambda}=0$, which completes the proof.

Let us introduce the notation%
\begin{equation}
\Lambda=\hbar\lambda,\ g\left(  \xi\right)  =\frac{e}{c\hbar}\mathcal{G}%
\left(  \xi\right)  ,\; p\left(  \xi\right)  =\lambda-g\left(  \xi\right)
,\ m=\frac{m_{0}c}{\hbar}~. \label{13}%
\end{equation}
Then (\ref{10}) can be rewritten as
\begin{equation}
P\,^{0}-P_{z}=\hbar p\left(  \xi\right)  ~, \label{14}%
\end{equation}
and (\ref{12}) implies%
\begin{equation}
m\dot{\xi}=p\left(  \xi\right)  \Longrightarrow s=\int\frac{md\xi}{p\left(
\xi\right)  }~, \label{15}%
\end{equation}
which relates the proper time and the parameter $\xi$.

In what follows, we denote%
\begin{equation}
r_{ij}\left(  \xi\right)  =r_{ji}\left(  \xi\right)  =\frac{e}{c\hbar}%
R_{ij}\left(  \xi\right)  ,\; F_{i}\left(  \xi\right)  =\frac{e}{c\hbar
}\mathcal{F}_{i}\left(  \xi\right)  ,\;\left(  i,j=1,2\right)  ,\; H\left(
\xi\right)  =\frac{e}{c\hbar}\mathcal{H}\left(  \xi\right)  ,\; S=\frac
{1}{\hbar}\mathcal{S}\,. \label{16}%
\end{equation}
Let us also introduce a $2\times2$ symmetric matrix $r=r\left(  \xi\right)  $
and the two-dimensional columns $F=F\left(  \xi\right)  $ and $v$,%
\begin{equation}
r=\left(
\begin{array}
[c]{cc}%
r_{11}\left(  \xi\right)  & r_{12}\left(  \xi\right) \\
r_{12}\left(  \xi\right)  & r_{22}\left(  \xi\right)
\end{array}
\right)  ,\; F=\left(
\begin{array}
[c]{c}%
F_{1}\left(  \xi\right) \\
F_{2}\left(  \xi\right)
\end{array}
\right)  ,\; v=\left(
\begin{array}
[c]{c}%
x\\
y
\end{array}
\right)  ~. \label{17}%
\end{equation}

The complete integral of the Hamilton--Jacobi equations (\ref{5}) for the
fields (\ref{2a}) can be presented as%
\begin{equation}
S=-\frac{1}{2}\left[  \lambda\left(  x^{0}+z\right)  +\Gamma\right]
,\;\Gamma=v^{+}fv+\chi^{+}v+v^{+}\chi+F^{+}v+v^{+}F+\int\left(  \chi^{+}%
\chi+m^{2}\right)  p^{-1}\left(  \xi\right)  d\xi~, \label{18}%
\end{equation}
where the $2\times2$ real symmetric matrix $f=f\left(  \xi\right)  $,%
\[
f=\left(
\begin{array}
[c]{cc}%
f_{11}\left(  \xi\right)  & f_{12}\left(  \xi\right) \\
f_{12}\left(  \xi\right)  & f_{22}\left(  \xi\right)
\end{array}
\right)  ,
\]
and the real two-column $\chi=\chi\left(  \xi\right)  $ satisfy the equations
(see Appendix I)%
\begin{align}
&  p\left(  \xi\right)  \left[  f^{\prime}\left(  \xi\right)  +r\left(
\xi\right)  \right]  -\left[  f\left(  \xi\right)  +iH\left(  \xi\right)
\sigma_{2}\right]  \left[  f\left(  \xi\right)  -iH\left(  \xi\right)
\sigma_{2}\right]  =0~,\label{19}\\
&  p\left(  \xi\right)  \left[  \chi\,^{\prime}\left(  \xi\right)
+F\,^{\prime}\left(  \xi\right)  \right]  -\left[  f\left(  \xi\right)
+iH\left(  \xi\right)  \sigma_{2}\right]  \chi\left(  \xi\right)  =0~.
\label{20}%
\end{align}
Here, $\sigma_{2}$ is a Pauli matrix.

Using (\ref{19}) and (\ref{20}), we can see that the three independent
functions $f_{ij}\left(  \xi\right)  $ provide a solution to a set of three
first-order non-linear equations, while the two functions $\chi_{i}\left(
\xi\right)  $ obey a set of two linear first-order inhomogenous equations,
where the functions $f_{ij}\left(  \xi\right)  $ are assumed to be known. One
should look for a particular solution of equations (\ref{19}), and so the
general solution of (\ref{20}) has the structure
\begin{equation}
\chi\left(  \xi\right)  =k_{1}\chi^{\left(  1\right)  }\left(  \xi\right)
+k_{2}\chi^{\left(  2\right)  }\left(  \xi\right)  +\bar{\chi}\left(
\xi\right)  ~. \label{21}%
\end{equation}
Here, $\bar{\chi}\left(  \xi\right)  $ is a particular solution for the set of
inhomogenous equations (\ref{20}); $\chi^{\left(  1\right)  }\left(
\xi\right)  $ and $\chi^{\left(  2\right)  }\left(  \xi\right)  $ provide a
fundamental system of solutions for the set of homogeneous equations
(\ref{20}); $k_{1}$ and $k_{2}$ are arbitrary constants (two integrals of
motion). Thus, the complete integral (\ref{18}) of the Hamilton--Jacobi
equations (\ref{5}) depends on three integrals of motion, $\lambda$, $k_{1}$
and $k_{2}$.

Having at one's disposal solutions of the equations (\ref{19}) and (\ref{20}),
one can easily find first integrals of the Lorentz equations. Using (\ref{4}),
with allowance made for (\ref{15}), one readily obtains a set of equations for
the coordinates $x$ and $y$ as functions of the variable $\xi$ in the
following matrix form:%
\begin{equation}
p\left(  \xi\right)  v^{\prime\prime}\left(  \xi\right)  +p^{\prime}\left(
\xi\right)  v^{\prime}\left(  \xi\right)  -\left[  r\left(  \xi\right)
+iH^{\prime}\left(  \xi\right)  \sigma_{2}\right]  v\left(  \xi\right)
-2iH\left(  \xi\right)  v^{\prime}\left(  \xi\right)  -F^{\prime}\left(
\xi\right)  =0~. \label{22}%
\end{equation}
One can easily prove that this equation can be integrated once,
\begin{equation}
p\left(  \xi\right)  v^{\prime}\left(  \xi\right)  +\left[  f\left(
\xi\right)  -iH\left(  \xi\right)  \sigma_{2}\right]  v\left(  \xi\right)
+\chi\left(  \xi\right)  =0 \label{23}%
\end{equation}
(see Appendix II).

Using identity (\ref{6}) for the kinetic momenta, relations (\ref{14}) and
(\ref{16}), we find%
\begin{align*}
&  ~(P\,^{0}-P_{z})(P\,^{0}+P_{z})=m_{0}^{2}c^{2}+P_{x}^{2}+P_{y}%
^{2}\Longrightarrow(P\,^{0}-P_{z})(P\,^{0}-P_{z}+2P_{z})\\
&  ~=m_{0}^{2}c^{2}+P_{x}^{2}+P_{y}^{2}\Longrightarrow p^{2}\left(
\xi\right)  \left[  1+2z\,^{\prime}\left(  \xi\right)  \right]  =m^{2}%
+p^{2}\left(  \xi\right)  v^{\prime}{}^{+}\left(  \xi\right)  v^{\prime
}\left(  \xi\right)  ~,
\end{align*}
the coordinate $z$ being a function of the variable $\xi$. We finally obtain%
\begin{equation}
2z\,^{\prime}\left(  \xi\right)  -m^{2}p^{-2}\left(  \xi\right)  -v^{\prime}%
{}^{+}\left(  \xi\right)  v^{\prime}\left(  \xi\right)  +1=0~. \label{24}%
\end{equation}
Expressions (\ref{23}) and (\ref{24}) are first integrals of the Lorentz equations.

\section{Crossed fields and beam guides}

In this section we study a particular case of the field (\ref{2a}) in the
absence of the longitudinal field, i.e., pure crossed-fields, and discuss how
these kinds of fields can be used to create a beam guide for charged
particles, i.e., fields that limit the motion of the charge around some given
trajectories. These guides trap the charge in a bidimensional plane
perpendicular to its trajectory and they are commonly used in many practical
applications, e.g., quantum computation \cite{PacWa02}, high resolution
spectroscopy \cite{BarMaHGK04}, non-neutral plasma physics \cite{HorDr02}, and
mass spectroscopy \cite{Gho97}. We will demonstrate that the Lorentz equations
for these crossed-fields can be reduced to the classical Newton equation with
a bidimensional effective potential. As an example we discuss a beam guide
created by an electromagnetic vortex \cite{Bir04}. Different from the
approximated classical analogues used to explain the operation of some RF
traps \cite{ThoHaB02}, the analysis developed here is exact and can be used to
describe the precise relativistic motion of the charge.

In the absence of the longitudinal field, we have%
\begin{equation}
E_{z}=\mathcal{G}^{\prime}\left(  \xi\right)  =H_{z}=\mathcal{H}\left(
\xi\right)  =0~. \label{m1}%
\end{equation}
As the only influence of the constant $\mathcal{G}$ manifests itself through
the $z$-component of the electric field, we can set $\mathcal{G}=0$. So our
potential (\ref{1a}) takes the form%
\begin{align}
&  A^{0}=A_{z}=-\frac{1}{2}A,\; A_{x}=-\mathcal{F}_{1}\left(  \xi\right)  ,\;
A_{y}=-\mathcal{F}_{2}\left(  \xi\right)  ~,\nonumber\\
&  A={R}_{11}\left(  \xi\right)  x^{2}+2{R}_{12}\left(  \xi\right)
xy+{R}_{22}\left(  \xi\right)  y^{2},\;\xi=x^{0}-z~. \label{m2}%
\end{align}
Whence, the fields%
\begin{align}
&  E_{x}=H_{y}=\mathcal{F}_{1}^{\prime}\left(  \xi\right)  +{R}_{11}\left(
\xi\right)  x+{R}_{12}\left(  \xi\right)  y~,\nonumber\\
&  E_{y}=-H_{x}=\mathcal{F}_{2}^{\prime}\left(  \xi\right)  +{R}_{12}\left(
\xi\right)  x+{R}_{22}\left(  \xi\right)  y~. \label{m2a}%
\end{align}
For these fields we can identify the integral of motion (\ref{10}) with the
light-front energy $E=\lambda/m$, and the parameter $\xi$ is directly
proportional to the proper time $s$, (\ref{15}) $s=m\xi/\lambda$, with
$\lambda$ and $m$ given by (\ref{13}). Substituting the fields (\ref{m2a}) in
the Lorentz equation (\ref{4}) we obtain:%
\begin{equation}
m\mathbf{\dot{P}}_{\perp}=\lambda\mathbf{P}_{\perp}^{\prime}=e\lambda
\mathbf{E}_{\perp}=-e\lambda\left(  \boldsymbol{\nabla} _{\perp}A_{0}%
+\partial_{0}\mathbf{A}_{\perp}\right)  \label{m3}%
\end{equation}
where the symbol $\perp$ indicates the perpendicular $x$ and $y$ components of
the vectors, e.g.,$\ \boldsymbol{\nabla} _{\perp}=\left(  \partial
_{x},\partial_{y}\right)  $. We can eliminate the perpendicular components
$\mathbf{A}_{\perp}$ of the potential (\ref{m2}) using the gauge
transformation%
\begin{align*}
&  \tilde{A}_{\mu}\left(  \xi,x,y\right)  =A_{\mu}\left(  \xi,x,y\right)
+\partial_{\mu}\phi\left(  \xi,x,y\right)  ~,\\
&  \phi\left(  \xi,x,y\right)  =x\mathcal{F}_{1}\left(  \xi\right)
+y\mathcal{F}_{2}\left(  \xi\right)  ~.
\end{align*}
So, making use of $\mathbf{P}_{\perp}^{\prime}=\hbar\lambda\mathbf{x}_{\perp
}^{\prime\prime}$, the equation (\ref{m3}) becomes%
\begin{equation}
\lambda\mathbf{x}_{\perp}^{\prime\prime}=-\frac{e}{\hbar}\boldsymbol{\nabla}
_{\perp}\tilde{A}_{0}~. \label{m3a}%
\end{equation}

We can identify the above expression with Newton's non-relativistic equation
for the two-dimensional motion of a particle with effective mass $\lambda$
moving in the effective potential%
\begin{equation}
U\left(  \xi,x,y\right)  =\frac{e}{\hbar}\tilde{A}_{0}=\frac{e}{\hbar}\left[
x\mathcal{F}_{1}^{\prime}\left(  \xi\right)  +y\mathcal{F}_{2}^{\prime}\left(
\xi\right)  -\frac{1}{2}A\left(  \xi,x,y\right)  \right]  ~. \label{m4}%
\end{equation}
Therefore, we can find fields that trap a charge in some point of the
$x,y$-plane without explicitly solving the Lorentz equations, just by looking
for functions $A,\mathcal{F}_{1}$ and $\mathcal{F}_{2}$ for which the
associated potential $U$ is capable of limiting the classical motion of a
particle of mass $\lambda$ around this point.

For the special case of a \textit{plane-wave}, where $\mathbf{E}%
=\mathbf{E}\left(  \xi\right)  $ and $\mathbf{H}=\mathbf{H}\left(  \xi\right)
$, we have $A=0$, which generates the effective potential%
\begin{equation}
U\left(  \xi,x,y\right)  =\frac{e}{\hbar}\left[  x\mathcal{F}_{1}^{\prime
}\left(  \xi\right)  +y\mathcal{F}_{2}^{\prime}\left(  \xi\right)  \right]  ~,
\label{m5}%
\end{equation}
and consequently creates a force $\hbar\mathbf{F}_{\perp}=e\left(
\mathcal{F}_{1}^{\prime},\mathcal{F}_{2}^{\prime}\right)  $ that does not
depend on the $x,y$ coordinates. So although a plane-wave may limit the motion
of a charge around some point, it is not possible to fix the position of this
point only by manipulating the fields.

\subsection{Time-independent fields}

For a \textit{time-independent potential}, the boundary trajectories can be
found by looking for the minima of the surface $U\left(  x,y\right)  $. These
points can be found using the standard procedure to determine the maxima and
minima of a function of several variables \cite{Buc78}. A point $\left(
x_{0},y_{0}\right)  $ will be an extreme if the first derivatives $\partial
U/\partial x$ and $\partial U/\partial y$\ vanish at this point, and this
extreme will be a minimum if the second derivative $\partial^{2}U/\partial
x^{2}$ and the discriminant $D\left(  x,y\right)  $ are positive at $\left(
x_{0},y_{0}\right)  $,%
\begin{equation}
D\left(  x_{0},y_{0}\right)  =\left(  \frac{\partial^{2}U}{\partial x^{2}%
}\frac{\partial^{2}U}{\partial y^{2}}\right)  -\left(  \frac{\partial^{2}%
U}{\partial x\partial y}\right)  ^{2}>0~,\ \left.  \frac{\partial^{2}%
U}{\partial x^{2}}\right\vert _{x_{0},y_{0}}>0~. \label{m5d}%
\end{equation}

In the case of a time-independent potential, the expression (\ref{m4}) for $U$
assumes the form%
\begin{equation}
U\left(  x,y\right)  =\frac{e}{\hbar}\left[  xC_{1}+yC_{2}-\frac{1}{2}\left(
x^{2}{R}_{11}+2xy{R}_{12}+y^{2}{R}_{22}\right)  \right]  ~, \label{m5e}%
\end{equation}
where $C_{i}$ and $R_{ij}$ $\left(  i,j=1,2\right)  $ are constants. So the
condition (\ref{m5d}) implies%
\begin{equation}
{R}_{11}<0\ \text{and}\ {R}_{11}{R}_{22}>\left(  {R}_{12}\right)  ^{2}~,
\label{m5f}%
\end{equation}
and, consequently, ${R}_{22}<0$ and $\det R\neq0$. Therefore, the minimum for
the potential (\ref{m5e}), under the above restrictions (which is the unique
extreme point of $U$ and, consequently, a global minimum), is the point%
\[
x_{0}=\frac{R_{22}C_{1}-R_{12}C_{2}}{\det R}~,y_{0}=\frac{R_{11}C_{2}%
-R_{12}C_{1}}{\det R}~.
\]
However, the fields (\ref{m2a}) associated to the potential (\ref{m5e}),
\begin{align*}
E_{x}  &  =H_{y}=C_{1}-x\left|  {R}_{11}\right|  -y\left|  {R}_{12}\right|
~,\\
E_{y}  &  =-H_{x}=C_{2}-x\left|  {R}_{12}\right|  -y\left|  {R}_{22}\right|
~,
\end{align*}
correspond to a problem of a constant charge density in the $x,y$\ plane and a
constant current density in the $z$ direction, which is nonrealistic.

\subsection{Periodic time-dependent fields}

For a \textit{periodic time-dependent potential}, and the special case of
\textit{linear} (or quasi-linear\footnote{That is, special systems for which
the non-linear terms can be neglected. see Chapter XIV of \cite{Gan59}.})
systems (which includes the case of quadratics potentials (\ref{m4})), where
we have%
\[
\boldsymbol{\nabla} _{\perp}\tilde{A}_{0}=M\left(  t\right)  \mathbf{x}%
_{\perp}~,
\]
with $M\left(  t\right)  $ a periodic time-dependent $2\times2$ matrix, the
stability of the potential (\ref{m4}) can be studied using the Lyapunov
criteria \cite{Gan59}. To use these criteria, we first substitute Newton's
second order equation (\ref{m3a}) by a pair of first order equations making
$\mathbf{u}_{\perp}=\mathbf{x}_{\perp}^{\prime}$, that turns (\ref{m3a})
equivalent to%
\begin{equation}
\mathbf{V}_{\perp}^{\prime}=\Xi\mathbf{V}_{\perp}~,\ \mathbf{V}_{\perp
}=\left(
\begin{array}
[c]{c}%
\mathbf{x}_{\perp}\\
\mathbf{u}_{\perp}%
\end{array}
\right)  ~,\ \Xi\left(  t\right)  =\left(
\begin{array}
[c]{cc}%
0 & I\\
M\left(  t\right)  & 0
\end{array}
\right)  ~. \label{m5a}%
\end{equation}
The motion is called stable around the point $\left(  0,0\right)  $\ if, for
every $\varepsilon>0$ we can find a $\delta>0$ such that for arbitrary
initials values $\mathbf{V}_{\perp}\left(  0\right)  $ with moduli less than
$\delta$ the coordinates $\mathbf{V}_{\perp}\left(  t\right)  $ remain of
moduli less than $\varepsilon$ for all the time $t\geq0$, i.e., the motion is
stable if%
\[
\vee\varepsilon>0,~\exists~\delta>0\therefore\left\vert \mathbf{V}_{\perp
}\left(  0\right)  \right\vert <\delta\Longrightarrow\left\vert \mathbf{V}%
_{\perp}\left(  t\right)  \right\vert <\varepsilon\ \left(  t\geq0\right)  ~.
\]
For periodic time-dependent $M\left(  t\right)  $ there always exists a
transformation $\mathcal{R}$ that leads to a static problem $\tilde{\Xi
}=\mathcal{R}\Xi\mathcal{R}^{-1}=\mathrm{const}$. having the same stability
character as $\Xi$ (see \cite{Gan59}, Vol. II, p.119). So, after applying this
transformation, the stability of the system can be analyzed by finding the
roots $\lambda_{k}$ of the characteristic polynomial%
\begin{equation}
\det\left(  \tilde{\Xi}-\lambda I\right)  =0~. \label{m5b}%
\end{equation}

The system is stable if:

\begin{enumerate}
\item $\operatorname{Re}\left(  \lambda_{k}\right)  \leq0$, for all
$\lambda_{k}$;

\item The pure imaginary characteristic values $\operatorname{Re}\left(
\lambda_{k}\right)  =0$ (if any such exist) are simple roots.
\end{enumerate}

If at least one of the above conditions is violated the system will be unstable.

Let us use the above procedure to analyze the fields generated by the
functions (\ref{36}) of the next section, for a pure crossed field ($H=0$). In
this case, we have%
\begin{align*}
&  {R}_{11}\left(  \xi\right)  =C_{1}+C_{2}\cos\omega\xi,~{R}_{22}\left(
\xi\right)  =C_{1}-C_{2}\cos\omega\xi~,\\
&  {R}_{12}\left(  \xi\right)  =C_{2}\sin\omega\xi\,,~\mathcal{F}%
_{1}=\mathcal{F}_{2}=0~,\;\omega,C_{1,2}=\mathrm{const}.
\end{align*}
Substituting these values in (\ref{m4}) we obtain the effective potential%
\begin{equation}
U\left(  \xi,x,y\right)  =-\frac{e}{2\hbar}\left[  \left(  x^{2}+y^{2}\right)
C_{1}+\left(  x^{2}-y^{2}\right)  C_{2}\cos\omega\xi+2xyC_{2}\sin\omega
\xi\right]  ~. \label{m7}%
\end{equation}
Changing to a rotating frame, that is, making the transformation,%
\begin{equation}
\mathbf{\tilde{x}}_{\perp}=R\mathbf{x}_{\perp}\,,\;R\left(  \xi\right)
=\left(
\begin{array}
[c]{cc}%
\cos\left(  \omega\xi/2\right)  & \sin\left(  \omega\xi/2\right) \\
-\sin\left(  \omega\xi/2\right)  & \cos\left(  \omega\xi/2\right)
\end{array}
\right)  \,, \label{m8}%
\end{equation}
the potential (\ref{m7}) becomes the following time-independent expression:%
\begin{equation}
\tilde{U}\left(  \tilde{x},\tilde{y}\right)  =-\frac{e}{2\hbar}\left(  \left(
\tilde{x}^{2}+\tilde{y}^{2}\right)  C_{1}+\left(  \tilde{x}^{2}-\tilde{y}%
^{2}\right)  C_{2}\right)  ~. \label{m9}%
\end{equation}
In the rotating frame the equation of motion (\ref{m3a}) becomes%
\begin{align}
&  \mathbf{\tilde{x}}_{\perp}^{\prime\prime}=\left(  RMR^{-1}-R\left(
R^{-1}\right)  ^{\prime\prime}\right)  \mathbf{\tilde{x}}_{\perp}-2R\left(
R^{-1}\right)  ^{\prime}\mathbf{\tilde{x}}_{\perp}^{\prime}~,\nonumber\\
&  M\left(  \xi\right)  =\frac{e}{\lambda\hbar}\left(
\begin{array}
[c]{cc}%
C_{1}+C_{2}\cos\left(  \omega\xi\right)  & C_{2}\sin\left(  \omega\xi\right)
\\
C_{2}\sin\left(  \omega\xi\right)  & C_{1}-C_{2}\cos\left(  \omega\xi\right)
\end{array}
\right)  \,, \label{m9a}%
\end{align}
and the expression (\ref{m5a}) assumes the form%
\[
\mathbf{\tilde{V}}_{\perp}^{\prime}=\tilde{\Xi}\mathbf{\tilde{V}}_{\perp
}\,,\;\mathbf{\tilde{V}}_{\perp}=\left(
\begin{array}
[c]{c}%
\mathbf{\tilde{x}}_{\perp}\\
\mathbf{\tilde{u}}_{\perp}%
\end{array}
\right)  ,\;\tilde{\Xi}=\left(
\begin{array}
[c]{cc}%
0 & \mathbf{I}\\
RMR^{-1}-R\left(  R^{-1}\right)  ^{\prime\prime} & -2R\left(  R^{-1}\right)
^{\prime}%
\end{array}
\right)  \,,
\]
where, from (\ref{m8}) and (\ref{m9a}), we see that $\tilde{\Xi}$ is the
constant matrix%
\[
\tilde{\Xi}=\left(
\begin{array}
[c]{cc}%
0 & \mathbf{I}\\
c_{1}+c_{2}\sigma_{3}+\omega^{2}/4 & i\sigma_{2}\omega
\end{array}
\right)  ,\;c_{i}=\frac{eC_{i}}{\lambda\hbar}~,
\]
where $\sigma_{i}$ are the Pauli matrices. The roots $\lambda_{k}$ of the
characteristic polynomial (\ref{m5b}) are%
\begin{align}
&  \lambda_{1}=-\lambda_{2}=\frac{1}{2}\sqrt{4c_{1}-\omega^{2}+4\sqrt
{c_{2}^{2}-\omega^{2}c_{1}}}~,\nonumber\\
&  \lambda_{3}=-\lambda_{4}=\frac{1}{2}\sqrt{4c_{1}-\omega^{2}-4\sqrt
{c_{2}^{2}\ -\omega^{2}c_{1}}~}. \label{m9b}%
\end{align}
Since each eigenvalue has a negative partner, the two Lyapunov conditions will
be satisfied only if $\operatorname{Re}\left(  \lambda_{k}\right)  =0$.

The vortex field analyzed in \cite{Bir04} is a special case of (\ref{m7}) for
$C_{1}=0$ and $C_{2}=B_{0}\omega$. In this case, the constant potential
(\ref{m9}) becomes%
\begin{equation}
\tilde{U}\left(  \tilde{x},\tilde{y}\right)  =-\frac{eB_{0}\omega}{2\hbar
}\left(  \tilde{x}^{2}-\tilde{y}^{2}\right)  ~. \label{m9c}%
\end{equation}
This potential describes the surface of a saddle that rotates (by \ref{m8}) in
the $x,y$ plane with angular velocity $\omega/2$ in time $\xi$ (or angular
velocity $\Omega/2=\omega\lambda/2m$ in the proper time $s$). The classical
motion of a particle in such a rotating-saddle potential is well known
\cite{ThoHaB02}. However, we were able to obtain some information about the
trajectories without really solving the equations of motion.

The condition $\operatorname{Re}\left(  \lambda_{k}\right)  =0$ for the
eigenvalues (\ref{m9b}) related to the potential (\ref{m9c}) determines the
expression%
\begin{equation}
\left\vert \omega\right\vert \geq\frac{4e}{\lambda\hbar}\left\vert
B_{0}\right\vert ~. \label{m10}%
\end{equation}
This inequality gives us a condition for which the potential (\ref{m9c})
generates bounded trajectories. The above result concurs with the condition
obtained in \cite{Bir04} by solving Lorentz's equation (\ref{4}) or the one
obtained in \cite{ThoHaB02} by solving the Newton's equation (\ref{m3a}).

\section{Solutions of Klein-Gordon and Dirac equations}

Solutions of the Klein--Gordon equation $\Phi_{\lambda,\,\mathbf{k}}\left(
\xi,\eta,x,y\right)  $ for the fields (\ref{2a}), labeled by the three
integrals of motion $\lambda$ (see \ref{13})) and $\mathbf{k}=\left(
k_{1},k_{2}\right)  $ (see (\ref{21})), read:
\begin{align}
&  \Phi_{\lambda,\,\mathbf{k}}\left(  \xi,\eta,x\, y\right)  =N_{0}%
p^{-1/2}\left(  \xi\right)  \sqrt{\Delta\left(  \xi\right)  }\exp\left(
iS\right)  ~,\nonumber\\
&  \Delta\left(  \xi\right)  =\det B\left(  \xi\right)  ,\; B\left(
\xi\right)  =\left(
\begin{array}
[c]{cc}%
\chi_{1}^{\left(  1\right)  }\left(  \xi\right)  & \chi_{1}^{\left(  2\right)
}\left(  \xi\right) \\
\chi_{2}^{\left(  1\right)  }\left(  \xi\right)  & \chi_{2}^{\left(  2\right)
}\left(  \xi\right)
\end{array}
\right)  ~,\nonumber\\
&  \eta=x^{0}+z=ct+z~, \label{25}%
\end{align}
where $N_{0}$ is a normalization factor and $\chi_{s\,^{\prime}}^{\left(
s\right)  }\left(  \xi\right)  \,\left(  s,\, s^{\prime}=1,2\right)  $ are the
spinor components $\chi^{\left(  s\right)  }\left(  \xi\right)  $ introduced
in (\ref{21}),%
\begin{equation}
\chi^{\left(  s\right)  }=\left(
\begin{array}
[c]{c}%
\chi_{1}^{\left(  s\right)  }\left(  \xi\right) \\
\chi_{2}^{\left(  s\right)  }\left(  \xi\right)
\end{array}
\right)  ,\; s=1,2~. \label{D3}%
\end{equation}
This fact can be directly verified by taking into account that the function
$\Delta\left(  \xi\right)  $ obeys the equation
\begin{equation}
p\frac{d\Delta}{d\xi}=\left(  \mathrm{tr}f\,\right)  \Delta\;, \label{26a}%
\end{equation}
which is a consequence of the uniform set (\ref{20}). Indeed, the spinors
$\chi^{\left(  s\right)  }$ obey the following equation:
\begin{equation}
\left(  \chi^{\left(  s\right)  }\right)  ^{\prime}=p^{-1}\left[
f+iH\sigma_{2}\right]  \chi^{\left(  s\right)  }\Longleftrightarrow\left(
\chi^{\left(  s\right)  +}\right)  ^{\prime}=p^{-1}\chi^{\left(  s\right)
+}\left[  f-iH\sigma_{2}\right]  ~. \label{D4}%
\end{equation}
The linear independence of the spinors $\chi^{\left(  s\right)  }$ implies
that the matrix $B$ from (\ref{25}) is nonsingular, i.e.,%
\begin{equation}
\Delta=\det B=\chi_{1}^{\left(  1\right)  }\left(  \xi\right)  \chi
_{2}^{\left(  2\right)  }\left(  \xi\right)  -\chi_{2}^{\left(  1\right)
}\left(  \xi\right)  \chi_{1}^{\left(  2\right)  }\left(  \xi\right)  \neq0~.
\label{D6}%
\end{equation}
The function $\Delta$ can be easily calculated. One can see that for real
spinors $\chi^{\left(  s\right)  }$ the following relations hold
\begin{equation}
\Delta=i\chi^{\left(  1\right)  +}\sigma_{2}\chi^{\left(  2\right)
}\Longrightarrow\Delta^{\prime}=i\left(  \chi^{\left(  1\right)  +}\right)
^{\prime}\sigma_{2}\chi^{\left(  2\right)  }+i\chi^{\left(  1\right)  +}%
\sigma_{2}\left(  \chi^{\left(  2\right)  }\right)  ^{\prime}~. \label{D7}%
\end{equation}
Then, using (\ref{D4}), we find
\[
\Delta^{\prime}=p^{-1}\chi^{\left(  1\right)  +}\left[  if\sigma_{2}%
+H+i\sigma_{2}f-H\right]  \chi^{\left(  2\right)  }=p^{-1}\chi^{\left(
1\right)  +}\left[  if\sigma_{2}+i\sigma_{2}f\right]  \chi^{\left(  2\right)
}~.
\]
With the help of an evident identity
\[
if\sigma_{2}+i\sigma_{2}f=\left(  \mathrm{tr}f\,\right)  i\sigma_{2}\,,
\]
we finally find (\ref{26a}).

Solutions of the Dirac equation $\Psi_{\lambda,\,\mathbf{k}}\left(  \xi
,\eta,x,y\right)  $ for the fields in question can be presented in a block
form by using the two-dimensional Pauli matrices:%
\begin{equation}
\Psi_{\lambda,\,\mathbf{k}}\left(  \xi,\eta,x,y\right)  =Np^{-1}\left(
\xi\right)  \sqrt{\Delta\left(  \xi\right)  }\exp\left[  iS\right]  \left(
\begin{array}
[c]{c}%
m+p\left(  \xi\right)  -\sigma_{3}\left(  \mathbf{\sigma\digamma}\right) \\
\left[  m-p\left(  \xi\right)  \right]  \sigma_{3}-\left(  \mathbf{\sigma
\digamma}\right)
\end{array}
\right)  V\left(  \xi\right)  ~. \label{26}%
\end{equation}
Here, the two-component spinor $V\left(  \xi\right)  $ reads%
\begin{equation}
V\left(  \xi\right)  =\left[  \cos T\left(  \xi\right)  +i\sigma_{3}\sin
T\left(  \xi\right)  \right]  V_{0}~, \label{27}%
\end{equation}
where $V_{0}$ is an arbitrary constant two-component spinor, and we also
denote%
\begin{equation}
T\left(  \xi\right)  =\int H\left(  \xi\right)  p^{-1}\left(  \xi\right)
~d\xi~. \label{28}%
\end{equation}
The components $\digamma_{i},$ $i=1,2,3$ of the vector $\mathbf{\digamma}$
have the form%
\[
\digamma_{1}=f_{11}\left(  \xi\right)  x+\left[  f_{12}\left(  \xi\right)
-H\left(  \xi\right)  \right]  y+\chi_{1}\left(  \xi\right)  ,\;\digamma
_{2}=\left[  f_{12}\left(  \xi\right)  +H\left(  \xi\right)  \right]
x+f_{22}\left(  \xi\right)  y+\chi_{1}\left(  \xi\right)  ,\;\digamma_{3}=0~.
\]

Therefore, the classical and quantum-mechanical problems are reduced to the
solution of the equations (\ref{19}) and (\ref{20}).

We will demonstrate that for a complete solution of the problem it is
sufficient to find a special particular solution of equations (\ref{19}).

The non-linear set of equations (\ref{19}) can be linearized by the following
substitution:
\begin{equation}
f\left(  \xi\right)  =p\left(  \xi\right)  \left[  \cos T\left(  \xi\right)
+i\sigma_{2}\sin T\left(  \xi\right)  \right]  Z^{\prime}\left(  \xi\right)
Z^{-1}\left(  \xi\right)  \left[  \cos T\left(  \xi\right)  -i\sigma_{2}\sin
T\left(  \xi\right)  \right]  ~, \label{29}%
\end{equation}
where $Z\left(  \xi\right)  $ is a non-degenerate second-order matrix. Using
(\ref{19}), we find a linear second-order equation for the matrix $Z\left(
\xi\right)  $,%
\begin{align}
&  p^{2}\left(  \xi\right)  Z^{\prime\prime}\left(  \xi\right)  +p\left(
\xi\right)  p^{\prime}\left(  \xi\right)  Z^{\prime}\left(  \xi\right)
+\left[  H^{2}\left(  \xi\right)  -p\left(  \xi\right)  \bar{r}\left(
\xi\right)  \right]  Z\left(  \xi\right)  =0~,\nonumber\\
&  \bar{r}\left(  \xi\right)  \equiv\left[  \cos T\left(  \xi\right)
-i\sigma_{2}\sin T\left(  \xi\right)  \right]  r\left(  \xi\right)  \left[
\cos T\left(  \xi\right)  +i\sigma_{2}\sin T\left(  \xi\right)  \right]  ~.
\label{30}%
\end{align}
A direct calculation yields%
\begin{align}
&  \bar{r}_{11}\left(  \xi\right)  =\frac{1}{2}r_{11}\left(  \xi\right)
\left[  1+\cos2T\left(  \xi\right)  \right]  +\frac{1}{2}r_{22}\left(
\xi\right)  \left[  1-\cos2T\left(  \xi\right)  \right]  -r_{12}\left(
\xi\right)  \sin2T\left(  \xi\right)  ~,\nonumber\\
&  \bar{r}_{12}\left(  \xi\right)  =\bar{r}_{21}\left(  \xi\right)
=r_{12}\left(  \xi\right)  \cos2T\left(  \xi\right)  +\frac{1}{2}\left[
r_{11}\left(  \xi\right)  -r_{22}\left(  \xi\right)  \right]  \sin2T\left(
\xi\right)  ~,\nonumber\\
&  \bar{r}_{22}\left(  \xi\right)  =\frac{1}{2}r_{22}\left(  \xi\right)
\left[  1+\cos2T\left(  \xi\right)  \right]  +\frac{1}{2}r_{11}\left(
\xi\right)  \left[  1-\cos2T\left(  \xi\right)  \right]  +r_{12}\left(
\xi\right)  \sin2T\left(  \xi\right)  ~. \label{31}%
\end{align}
In order that the matrix $f\left(  \xi\right)  $ be real and symmetric, one
has to look for real solutions of the equations (\ref{30}) that obey the
subsidiary condition (the symbol $\sim$ stands for transposition)%
\begin{equation}
J\left(  \xi\right)  =\tilde{J}\left(  \xi\right)  ,\; J\left(  \xi\right)
\equiv Z^{\prime}\left(  \xi\right)  Z^{-1}\left(  \xi\right)  ~. \label{32}%
\end{equation}
Such solutions of (\ref{30}) always exist, since, in accordance with
(\ref{30}), one easily finds an equation for $J\left(  \xi\right)  $,%
\begin{equation}
p^{2}\left(  \xi\right)  J^{\prime}\left(  \xi\right)  +p^{2}\left(
\xi\right)  J^{2}\left(  \xi\right)  +p\left(  \xi\right)  p^{\prime}\left(
\xi\right)  J\left(  \xi\right)  +H^{2}\left(  \xi\right)  -p\left(
\xi\right)  \bar{r}\left(  \xi\right)  =0~. \label{33}%
\end{equation}
Transposing this equation and using the property $\tilde{\bar{r}}\left(
\xi\right)  =\bar{r}\left(  \xi\right)  $, we find that the equations for
$J\left(  \xi\right)  $ and $\tilde{J}\left(  \xi\right)  $ are the same,
which proves the existence of solutions that satisfy (\ref{32}).

Having at one's disposal a particular solution of equations (\ref{30}) that
satisfies the condition (\ref{32}), one can integrate the equations (\ref{20})
and (\ref{23}) by quadratures. Indeed, a direct verification shows that the
expressions%
\begin{align}
&  v=\left[  \cos T\left(  \xi\right)  +i\sigma_{2}\sin T\left(  \xi\right)
\right]  Z\left(  \xi\right)  \left\{  v_{0}-\int Z^{-1}\left(  \xi\right)
\left[  \cos T\left(  \xi\right)  -i\sigma_{2}\sin T\left(  \xi\right)
\right]  \chi\left(  \xi\right)  p^{-1}\left(  \xi\right)  d\xi\right\}
~,\nonumber\\
&  \chi=\left[  \cos T\left(  \xi\right)  +i\sigma_{2}\sin T\left(
\xi\right)  \right]  \tilde{Z}^{-1}\left(  \xi\right)  \left\{  K-\int
\tilde{Z}\left(  \xi\right)  \left[  \cos T\left(  \xi\right)  -i\sigma
_{2}\sin T\left(  \xi\right)  \right]  F\,^{\prime}\left(  \xi\right)
d\xi\right\}  ~, \label{34}%
\end{align}
obey equations (\ref{20}) and (\ref{23}), respectively, where $K$ is a column
with components $k_{1}$ and $k_{2}$, and $v_{0}$ is a constant two-component spinor.

Motion in the fields of the type (\ref{2a}) has been studied in previous
works. The authors of \cite{1} found the symmetry operators for these fields,
and the authors of \cite{2,3,4,5} found solutions for numerous specific fields
of this kind. However, in all these specific fields there exist certain
transformations that diagonalize $\bar{r}\left(  \xi\right)  $, so that the
set of equations (\ref{30}) splits into independent linear equations of second order.

A considerable progress was made in the work \cite{Bir04}, which was the first
to present exact solutions for a specific field that does not admit any
transformations leading to the separation of equations (\ref{30}) into
independent equations of second order. The author of \cite{Bir04} examined a
particular case of the fields (\ref{2a}) with the following choice of
functions:%
\begin{equation}
\mathcal{F}_{i}\left(  \xi\right)  =H\left(  \xi\right)  =g\left(  \xi\right)
=0,\;{r}_{11}\left(  \xi\right)  =-{r}_{22}\left(  \xi\right)  =c\cos\omega
\xi,\;{r}_{12}\left(  \xi\right)  =c\sin\omega\xi~, \label{35}%
\end{equation}
where $\omega$ and $c$ are some constants. The solutions of the equations in
\cite{Bir04} were obtained in a different manner from that of the approach of
the present work, and they depend essentially on the specific form of the fields.

Let us note that in the present approach there is no necessity to assume
$\mathcal{F}_{i}\left(  \xi\right)  =0$, because these functions do not enter
the set of equations (\ref{30}), and so they may be left arbitrary.

We have succeeded in finding solutions for the fields defined by the following
functions:
\begin{align}
&  g\left(  \xi\right)  =0,\;\; H\left(  \xi\right)  =H=\mathrm{const}%
,\;\;{r}_{11}\left(  \xi\right)  =c_{1}+c_{2}\cos\omega\xi~,\nonumber\\
&  {r}_{22}\left(  \xi\right)  =c_{1}-c_{2}\cos\omega\xi,\;\;{r}_{12}\left(
\xi\right)  =c_{2}\sin\omega\xi,\;\;\omega,\, c_{1,2}=\mathrm{const~}.
\label{36}%
\end{align}
A peculiarity of (\ref{36}) is the presence of a constant and homogenous
magnetic field. Expressions (\ref{35}) provide a particular case of (\ref{36})
with $H=c_{1}=0$.

It is easy to see that in the case of the functions (\ref{36}), the equation
(\ref{30}) can be written in the form%
\begin{equation}
\lambda^{2}Z^{\prime\prime}+\left[  H^{2}-\lambda c_{1}-\lambda c_{2}\left(
\mathbf{\sigma l}\right)  \right]  Z=0,\quad\mathbf{l}=\left(  \sin\Omega
\,\xi,0,\cos\Omega\,\xi\right)  ~, \label{37}%
\end{equation}
where $T$ is%
\[
T\left(  \xi\right)  =\frac{H\xi}{\lambda},\;\Omega=\omega+\frac{2H}{\lambda
}~.
\]
For $c_{2}=0$, solutions of this equation are known \cite{2a,3,4,5}, and
therefore we only need to examine the case $c_{2}\neq0$. A direct verification
shows that the expressions%
\begin{align}
&  Z_{11}=A\left[  \alpha\cos\frac{\Omega\,x}{2}\cos\alpha\xi\,+\left(
\frac{\Omega}{2}+\frac{\gamma+c_{2}}{\lambda\,\Omega}\right)  \,\sin
\frac{\Omega\,x}{2}\sin\alpha\xi\right]  ~,\nonumber\\
&  Z_{21}=A\left[  \alpha\sin\frac{\Omega\,x}{2}\cos\alpha\xi-\left(
\frac{\Omega}{2}+\frac{\gamma+c_{2}}{\lambda\,\Omega}\right)  \cos\frac
{\Omega\,x}{2}\sin\alpha\xi\right]  ~,\nonumber\\
&  Z_{12}=B\left[  \beta\sin\frac{\Omega\,x}{2}\sin\beta\xi+\left(
\frac{\Omega}{2}-\frac{\gamma+c_{2}}{\lambda\,\Omega}\right)  \cos\frac
{\Omega\,x}{2}\cos\beta\xi\right]  ~,\nonumber\\
&  Z_{22}=-B\left[  \beta\cos\frac{\Omega\,x}{2}\sin\beta\xi-\left(
\frac{\Omega}{2}-\frac{\gamma+c_{2}}{\lambda\,\Omega}\right)  \sin\frac
{\Omega\,x}{2}\cos\beta\xi\right]  ~, \label{38}%
\end{align}
where%
\begin{align*}
&  \alpha^{2}=\frac{H^{2}-\lambda c_{1}}{\lambda^{2}}+\frac{\Omega^{2}}%
{4}+\frac{\gamma}{\lambda},\;\beta^{2}=\frac{H^{2}-\lambda c_{1}}{\lambda^{2}%
}+\frac{\Omega^{2}}{4}-\frac{\gamma}{\lambda}~,\\
&  \gamma^{2}=c_{2}^{2}+(H^{2}-\lambda c_{1})\Omega^{2}%
,\;\;A,\,B=\mathrm{const.}%
\end{align*}
give a solution for the equations (\ref{37}) that satisfies the condition
(\ref{32}). The signs of the quantities $\alpha,\,\beta,\,\gamma$ may be
chosen arbitrarily. The quantity $\gamma$ is either real or purely imaginary;
the quantities $\alpha,\,\beta$ may be complex. For complex $\alpha,\,\beta$,
in view of the linear character of equations (\ref{37}), the real and
imaginary parts of (\ref{38}) separately provide the sought solutions. The
expressions (\ref{38}) admit a continuous limiting process $\Omega
\rightarrow0$ in case the sign of $\gamma$ ($\gamma$ being real when
$\Omega\rightarrow0$) is chosen to obey the condition $c_{2}\gamma<0$. By
carrying out this limiting process (and redefining the constants $A,B$) we
find, due to (\ref{38}) and $\Omega=0$, that%
\begin{equation}
Z\left(  \xi\right)  =\left(
\begin{array}
[c]{cc}%
A\cos\alpha\xi & 0\\
0 & B\sin\beta\xi
\end{array}
\right)  ,\quad%
\begin{array}
[c]{c}%
\alpha^{2}=\left(  H^{2}-\lambda c_{1}-\lambda c_{2}\right)  \lambda^{-2}\\
\beta^{2}=\left(  H^{2}-\lambda c_{1}+\lambda c_{2}\right)  \lambda^{-2}%
\end{array}
~. \label{39}%
\end{equation}
The calculation of the quantities $f\left(  \xi\right)  ,\,\chi\left(
\xi\right)  ,\,v\left(  \xi\right)  $, with the help of formulas (\ref{29})
and (\ref{34}), is reduced to simple algebraic manipulations and integrations
of elementary functions, which we omit.

\section{Orthogonality and completeness relations}

In the case under consideration, it is convenient to define inner products for
both scalar and spinor wave functions on the null-plane $\xi=\mathrm{const},$
see for details \cite{6,31}. Such an inner product for scalar wave functions
is%
\begin{align}
&  \,(\Phi_{1},\,\Phi_{2})_{\xi}=\int\left\{  \left[  \hat{Q}\Phi_{1}\left(
\xi,\eta,x,y\right)  \right]  ^{\ast}\Phi_{2}\left(  \xi,\eta,x,y\right)
\right. \nonumber\\
&  \,\left.  +\Phi_{1}^{\ast}\left(  \xi,\eta,x,y\right)  \hat{Q}\Phi
_{2}\left(  \xi,\eta,x,y\right)  \right\}  ~d\eta dxdy~, \label{31a}%
\end{align}
where
\[
\hat{Q}=\frac{\hat{P}_{0}-\hat{P}_{z}}{\hbar}=2i\frac{\partial}{\partial\eta
}-g\left(  \xi\right)  ~.
\]
For spinor wave functions, the inner product on the null-plane has the form
\begin{align}
&  \,\langle\Psi_{1},\,\Psi_{2}\rangle_{\xi}=\int\Psi_{1\left(  -\right)
}^{+}\left(  \xi,\eta,x,y\right)  \Psi_{2(-)}\left(  \xi,\eta,x,y\right)
~d\eta dxdy~,\nonumber\\
&  \Psi_{\left(  -\right)  }\hat{=}P_{\left(  -\right)  }\Psi,\;\hat
{P}_{\left(  -\right)  }=\frac{1}{2}\left(  1-\alpha_{3}\right)  =\frac{1}%
{2}\left(
\begin{array}
[c]{cc}%
I & -\sigma_{3}\\
-\sigma_{3} & I
\end{array}
\right)  ~. \label{32a}%
\end{align}

One can verify that scalar wave functions (\ref{25}) obey the orthonormality
condition
\begin{equation}
\left(  \Phi_{\lambda^{\prime},\,\mathbf{k}^{\prime}},\,\Phi_{\lambda
,\,\mathbf{k}}\right)  _{\xi}=\varepsilon\,\delta\left(  \lambda^{\prime
}-\lambda\right)  \delta\left(  k_{1}^{\prime}-k_{1}\right)  \delta\left(
k_{2}^{\prime}-k_{2}\right)  ,\;\varepsilon=p\left(  \xi\right)  |p\left(
\xi\right)  |^{-1}~, \label{33a}%
\end{equation}
provided $N_{0}=\left(  32\pi^{3}\right)  ^{-1/2}$. First we note that the
following relation holds
\begin{equation}
\hat{Q}\,\Phi_{\lambda,\,\mathbf{k}}\left(  \xi,\eta,x,y\right)  =p\left(
\xi\right)  \,\Phi_{\lambda,\,\mathbf{k}}\left(  \xi,\eta,x,y\right)  ~.
\label{D9}%
\end{equation}
Then, integration over the variable $\eta$ is reduced to the calculation of
the integral
\begin{equation}
\int_{-\infty}^{\infty}\exp\left[  \frac{i}{2}\left(  \lambda^{\prime}%
-\lambda\right)  \eta\right]  ~d\eta=4\pi\delta\left(  \lambda^{\prime
}-\lambda\right)  ~. \label{D10}%
\end{equation}
Therefore, we can set $\lambda^{\prime}=\lambda$ in the integral over $x,\,
y$. The latter is reduced to the following integral
\begin{align}
&  J=\int_{-\infty}^{\infty}dx\int_{-\infty}^{\infty}dy~\exp\left[
ix\sum_{s=1,\,2}\chi_{1}^{\left(  s\right)  }\left(  \xi\right)  \left(
k_{s}^{\prime}-k_{s}\right)  +iy\sum_{s=1,\,2}\chi_{2}^{\left(  s\right)
}\left(  \xi\right)  \left(  k_{s}^{\prime}-k_{s}\right)  \right] \nonumber\\
&  \,=4\pi^{2}\delta\left(  \sum_{s=1,\,2}\chi_{1}^{\left(  s\right)  }\left(
\xi\right)  \left(  k_{s}^{\prime}-k_{s}\right)  \right)  \delta\left(
\sum_{s=1,\,2}\chi_{2}^{\left(  s\right)  }\left(  \xi\right)  \left(
k_{s}^{\prime}-k_{s}\right)  \right)  ~. \label{D11}%
\end{align}

The product of two $\delta$-functions in the right hand side of (\ref{D11})
can be transformed if we take into account the following fact: Let $a$ be a
nonsingular $2\times2$ matrix, $\det a\neq0$, with matrix elements $a_{ij}$ .
Then the relation holds
\begin{equation}
\delta\left(  a_{11}z_{1}+a_{12}z_{2}\right)  \delta\left(  a_{21}z_{1}%
+a_{22}z_{2}\right)  =|\det a|^{-1}\delta\left(  z_{1}\right)  \delta\left(
z_{2}\right)  ~. \label{D2}%
\end{equation}

Setting in (\ref{D2}): $a_{ij}=\chi_{i}^{\left(  j\right)  }\left(
\xi\right)  ,\;\, z_{1}=k_{1}^{\prime}-k_{1},\;\, z_{2}=k_{2}^{\prime}-k_{2}$,
we obtain
\begin{equation}
J=4\pi^{2}|\Delta\left(  \xi\right)  |^{-1}\delta\left(  k_{1}^{\prime}%
-k_{1}\right)  \delta\left(  k_{2}^{\prime}-k_{2}\right)  ~, \label{D12}%
\end{equation}
where $\Delta\left(  \xi\right)  $ is given by (\ref{D6}). Then the result
(\ref{33a}) follows.

The constant spinor $V_{0}$ in solutions (\ref{27}) is related to an
additional (spinning) integral of motion, see for details \cite{5,6}. So the
spinor $V_{0}$ (and therefore the Dirac wave function as well) depends on a
spinning quantum number $\zeta=\pm1,\ V_{0}=V_{0}\left(  \zeta\right)  $. It
is always possible to choose $V_{0}\left(  \zeta\right)  $ such that it obeys
the following relations of orthonormality and completeness:%
\begin{equation}
V_{0}^{+}\left(  \zeta^{\prime}\right)  V_{0}\left(  \zeta\right)
=\delta\,_{\zeta,\,\zeta\,^{\prime}}\,;\;\sum_{\zeta=\pm1}V_{0}\left(
\zeta\right)  V_{0}^{+}\left(  \zeta\right)  =I~. \label{29a}%
\end{equation}

Taking into account (\ref{29a}) and the relation%
\begin{equation}
\Psi_{\left(  -\right)  \,\lambda,\,\mathbf{k},\,\zeta}\left(  \xi
,\eta,x,y\right)  =(32\pi^{3})^{-1/2}\Delta^{1/2}\left(  \xi\right)
\exp\left(  iS\right)  \left(
\begin{array}
[c]{c}%
I\\
-\sigma_{3}%
\end{array}
\right)  V\left(  \xi\right)  ~, \label{34a}%
\end{equation}
we can verify that the spinor wave functions (\ref{27}) obey the
orthonormality condition%
\begin{equation}
\langle\Psi_{\lambda^{\prime},\,\mathbf{k}^{\prime},\,\zeta\,^{\prime}}%
,\,\Psi_{\lambda,\,\mathbf{k},\,\zeta\,}\rangle_{\xi}=\delta\left(
\lambda^{\prime}-\lambda\right)  \delta\left(  k_{1}^{\prime}-k_{1}\right)
\delta\left(  k_{2}^{\prime}-k_{2}\right)  \delta_{\zeta,\,\zeta\,^{\prime}}
\label{35a}%
\end{equation}
provided $N=(32\pi^{3})^{-1/2}$.

The solutions (\ref{25}) and (\ref{27}) form complete sets of functions on the
null-plane $\xi=\mathrm{const}.$

For scalar wave functions (\ref{25}), we consider the following integral:%
\begin{equation}
M=2\int_{-\infty}^{\infty}d\lambda\int_{-\infty}^{\infty}dk_{1}\int_{-\infty
}^{\infty}dk_{2}~|p\left(  \xi\right)  |\,\Phi_{\lambda,\,\mathbf{k}}^{\ast
}(\xi,\,\eta^{\prime},\, x^{\prime},\, y^{\prime})\,\Phi_{\lambda
,\,\mathbf{k}}\left(  \xi,\eta,x,y\right)  ~. \label{D13}%
\end{equation}
Integrating over the variables $k_{1},\, k_{2}$ leads us to the integral:
\begin{align*}
&  M_{1}=\int_{-\infty}^{\infty}dk_{1}\int_{-\infty}^{\infty}dk_{2}%
~\exp\left(  iR^{\left(  1\right)  }k_{1}+iR^{\left(  2\right)  }k_{2}\right)
=4\pi^{2}\delta\left(  R^{\left(  1\right)  }\right)  \delta\left(  R^{\left(
2\right)  }\right)  ~,\\
&  R^{\left(  s\right)  }=\chi_{1}^{\left(  s\right)  }\left(  \xi\right)
\left(  x^{\prime}-x\right)  +\chi_{2}^{\left(  s\right)  }\left(  \xi\right)
\left(  y^{\prime}-y\right)  ,\; s=1,\,2~.
\end{align*}
This expression has the form (\ref{D2}), where $a=\tilde{B}\left(  \xi\right)
$ is the transpose of the matrix $B$ in (\ref{25}), and $z_{1}=x^{\prime}-x,\,
z_{2}=y^{\prime}-y$. Thus we obtain:
\begin{equation}
M_{1}=4\pi^{2}|\Delta\left(  \xi\right)  |^{-1}\delta\left(  x^{\prime
}-x\right)  \delta\left(  y^{\prime}-y\right)  ~. \label{D25}%
\end{equation}
After that one can easily integrate over $\lambda$ in (\ref{D13}) to get the
following completeness relation:
\begin{align*}
&  2\int_{-\infty}^{\infty}d\lambda\int_{-\infty}^{\infty}dk_{1}\int_{-\infty
}^{\infty}dk_{2}~|p\left(  \xi\right)  |\,\Phi_{\lambda,\,\mathbf{k}}^{\ast
}\left(  \xi,\eta^{\prime},x^{\prime},y^{\prime}\right)  \,\Phi_{\lambda
,\,\mathbf{k}}\left(  \xi,\eta,x,y\right) \\
&  \,=\delta\left(  x^{\prime}-x\right)  \delta\left(  y^{\prime}-y\right)
\delta\left(  \eta^{\prime}-\eta\right)  ~.
\end{align*}

Similar calculations can be performed in the spinor case. Here we have
additionally to use the second relation (\ref{29a}) to get a completeness
relation for the solutions (\ref{27}):
\begin{align*}
&  \sum_{\zeta=\pm1}\int_{-\infty}^{\infty}d\lambda\int_{-\infty}^{\infty
}dk_{1}\int_{-\infty}^{\infty}dk_{2}~\Psi_{\left(  -\right)  \,\lambda
,\,\mathbf{k},\,\zeta}^{+}\left(  \xi,\eta^{\prime},x^{\prime},y^{\prime
}\right)  \,\Psi_{\left(  -\right)  \,\lambda,\,\mathbf{k},\,\zeta}\left(
\xi,\eta,x,y\right) \\
&  \,=\delta\left(  x^{\prime}-x\right)  \delta\left(  y^{\prime}-y\right)
\delta\left(  \eta^{\prime}-\eta\right)  \hat{P}_{\left(  -\right)  }~.
\end{align*}

\begin{acknowledgement}
This work was partially supported by RFBR grant 06-02-16719 and Russia
President grant SS-5103.2006.2; M.C.B. thanks FAPESP; D.M.G. thanks FAPESP and
CNPq for permanent support.
\end{acknowledgement}

\renewcommand{\thesection}{}

\section{Appendix}

\textbf{I.} Equations (\ref{19}) and (\ref{20}) are obtained as follows. We
search for a complete integral of the Hamilton--Jacobi equations (\ref{5}) in
the form%
\begin{equation}
S=-\frac{1}{2}\left[  \lambda\left(  x^{0}+z\right)  +\Gamma\right]  ~,
\label{A1}%
\end{equation}
with the function $\Gamma$ is%
\begin{equation}
\Gamma=f_{11}\left(  \xi\right)  x^{2}+2f_{12}\left(  \xi\right)
xy+f_{22}\left(  \xi\right)  y^{2}+2\left[  \chi_{1}\left(  \xi\right)
+F_{1}\left(  \xi\right)  \right]  x+2\left[  \chi_{2}\left(  \xi\right)
+F_{2}\left(  \xi\right)  \right]  y+\alpha\left(  \xi\right)  ~. \label{A2}%
\end{equation}
Here, $f_{ij}\left(  \xi\right)  ,\,\chi_{i}\left(  \xi\right)  $, and
$\alpha\left(  \xi\right)  $ are unknown functions of the variable $\xi$.
Substituting the expression (\ref{A1}), with allowance made for (\ref{A2}),
into equation (\ref{5}), we obtain a quadratic form in $x,y$, with
coefficients being functions of $\xi$, that must be identically zero, which is
only possible when each coefficient is equal to zero. Hence, we obtain the
following equations:%
\begin{align}
&  p\left(  f_{11}^{\prime}+r_{11}\right)  -f_{11}^{2}-\left(  f_{12}%
+H\right)  ^{2}=0~,\label{A3}\\
&  p\left(  f_{22}^{\prime}+r_{22}\right)  -f_{22}^{2}-\left(  f_{12}%
-H\right)  ^{2}=0~,\label{A4}\\
&  p\left(  f_{12}^{\prime}+r_{12}\right)  -f_{11}\left(  f_{12}-H\right)
-f_{22}\left(  f_{12}+H\right)  =0~,\label{A5}\\
&  p\left(  \chi_{1}^{\prime}+F\,_{1}^{\prime}\right)  -f_{11}\chi_{1}-\left(
f_{12}+H\right)  \chi_{2}=0~,\label{A6}\\
&  p\left(  \chi_{2}^{\prime}+F\,_{2}^{\prime}\right)  -f_{22}\chi_{2}-\left(
f_{12}-H\right)  \chi_{1}=0~,\label{A7}\\
&  p\,\alpha\,^{\prime}-\chi_{1}^{2}-\chi_{2}^{2}-m^{2}=0~. \label{A8}%
\end{align}
Owing to (\ref{A8}), the expressions (\ref{18}) and (\ref{A2}) for the
function $\Gamma$ are identical. It is easy to see that the set of equations
(\ref{A3})--(\ref{A5}) coincides with the matrix equation (\ref{19}), and the
set of equations (\ref{A6})--(\ref{A7}) has the matrix form (\ref{20}).

\textbf{II.} Let us show that the equation (\ref{22}) is a consequence of the
equation (\ref{23}). To this end, we differentiate (\ref{23}) with respect to
the variable $\xi$,%
\begin{equation}
pv^{\prime\prime}+p\,^{\prime}v^{\prime}+\left[  f^{\prime}-iH\,^{\prime
}\sigma_{2}\right]  v+\left[  f-iH\sigma_{2}\right]  v^{\prime}+\chi^{\prime
}=0~. \label{B1}%
\end{equation}
However, the left-hand side of (\ref{B1}) satisfies the identity%
\begin{align}
&  pv^{\prime\prime}+p\,^{\prime}v^{\prime}+\left[  f^{\prime}-iH\,^{\prime
}\sigma_{2}\right]  v+\left[  f-iH\sigma_{2}\right]  v^{\prime}+\chi^{\prime
}=A+B~,\nonumber\\
&  A=pv^{\prime\prime}+p^{\prime}v^{\prime}-\left[  r+iH\,^{\prime}\sigma
_{2}\right]  v-2iH\sigma_{2}v^{\prime}~,\nonumber\\
&  B=\left[  f^{\prime}+r\right]  v+\left[  f+iH\sigma_{2}\right]  v^{\prime
}+\chi^{\prime}~. \label{B2}%
\end{align}
In the expression for $B$ we now substitute $f^{\prime}$ from (\ref{19}),
$\chi^{\prime}$ from (\ref{20}), and $v^{\prime}$ from (\ref{23}), and thus we
easily find that $B=-F^{\prime}$ and that $A+B$ is identical to the left-hand
side of (\ref{22}), which completes the proof.

\end{document}